\documentclass[floatfix,aps,prd,abstract,10pt,twocolumn,nofootinbib,preprintnumbers]{revtex4-1}

\usepackage{amsgen,amsmath,amstext,amsbsy,amsopn,amssymb}
\usepackage{graphicx}
\usepackage[usenames]{color}
\usepackage{epsfig}
\usepackage{multirow}
\usepackage{longtable}
\usepackage{bm} 
\usepackage{hyperref}
\usepackage{multirow}

\newcommand{\Ddel}{\delta_{\rm D}   }

\newcommand{\MpcOh}{ \,  \mathrm{Mpc}  \, h^{-1} }
\newcommand{\hOMpc}{ \,  \mathrm{Mpc}^{-1}  \, h  }

\newcommand{\hOMpcCube}{ \,  \mathrm{Mpc}^{-3}  \, h^3  }

\newcommand{\comment}[1]{}

\newcommand{\dv}{ \delta_{\rm v}}
\newcommand{\dc}{ \delta_{\rm c} }
\newcommand{\Msun}{ \,   M_{\odot}  {h}^{-1}   }
\newcommand{\fnl}{ f_{\rm NL} }

\newcommand{\beq}{\begin{equation}}
\newcommand{\eeq}{\end{equation}}
\newcommand{\beqa}{\begin{eqnarray}}
\newcommand{\eeqa}{\end{eqnarray}}

\begin{document}

\title{  Constraint of Void Bias on Primordial non-Gaussianity }

\author{Kwan Chuen Chan$^{(1)}$} \email{chankc@mail.sysu.edu.cn}
\author{Nico Hamaus$^{(2)} $}
\author{Matteo Biagetti$^{(3)}$}
\affiliation{$^{(1)}$ School of Physics and Astronomy, Sun Yat-Sen University, Guangzhou 510275, China }
\affiliation{$^{(2)}$ Universit\"ats-Sternwarte M\"unchen, Fakult\"at f\"ur Physik, Ludwig-Maximilians Universit\"at, Scheinerstr.~1, 81679 M\"unchen, Germany } 
\affiliation{$^{(3)}$ Institute of Physics, Universiteit van Amsterdam, Science Park, 1098XH Amsterdam,  Netherlands }

\date{\today}

\begin{abstract}
We study the large-scale bias parameter of cosmic voids with primordial non-Gaussian (PNG) initial conditions of the local type. In this scenario, the dark matter halo bias exhibits a characteristic scale dependence on large scales, which  has been recognized as one of the most promising probes of the local PNG. Using a suite of  $N$-body simulations with Gaussian and non-Gaussian initial conditions, we find that the void bias features scale-dependent corrections on large scales, similar to its halo counterpart. We find excellent agreement between the numerical measurement of the PNG void bias and the general peak-background split prediction. Contrary to halos, large voids anti-correlate with the dark matter density field, and the large-scale Gaussian void bias ranges from positive  to negative values depending on void size and redshift. Thus, the information in the clustering of voids can be complementary to that of the halos. Using the Fisher matrix formalism for multiple tracers, we demonstrate that including the scale-dependent bias information from voids, constraints on the PNG parameter $\fnl$ can be tightened by a factor of two compared to the accessible information from halos alone, when the sampling density of tracers reaches  $4 \times 10^{-3} \, \hOMpcCube $.
\end{abstract}

\maketitle

\section{Introduction}
Primordial non-Gaussianity (PNG) offers an important probe into the physics of inflation \cite{Bartolo:2004if,Liguori_etal2010,Chen:2010xka}, as it allows to constrain the production mechanism of primordial perturbations which seed the structures we observe in the Universe today. Furthermore, PNG can be directly related to primordial interactions taking place at energies as high as $10^{14}$ GeV, providing a unique window into the particle content of inflation \cite{Chen:2010xka,Arkani-Hamed:2015bza}.  In the local PNG model, the Bardeen potential $\Phi$ is given by \cite{Gangui:1993tt,Verde:1999ij,Komatsu:2001rj}
\beq
\label{label:Phi_localPNG}
\Phi = \phi + \fnl ( \phi^2 - \langle  \phi^2 \rangle  ), 
\eeq
where $\phi  $ is a Gaussian potential field and  $\fnl$ parametrizes the strength of non-Gaussianity.  Using maps of the CMB, the Planck collaboration derived a stringent constraint of $ \fnl = 0.8 \pm 5.0$  \cite{Ade:2015ava}.

Besides the CMB, the large-scale structure (LSS) is another frontier in constraining PNG. It can be detected using the galaxy bispectrum \cite{Verde:1999ij,Scoccimarro:2003wn,Sefusatti:2007ih}. However, to extract the feeble PNG signal from it, dominant contributions from late-time non-Gaussianities due to dark matter nonlinearities, galaxy biasing, and redshift-space distortions must be modeled well. Recovering information from the bispectrum is also hampered by its large covariance \cite{Chan:2016ehg}. On the other hand, it has been discovered that the halo bias exhibits a strong scale dependence on large scales in the local PNG scenario~\cite{Dalal:2007cu}. In contrast, for Gaussian initial conditions the large-scale halo bias remains scale-independent. Scale-dependent halo bias from PNG has been extensively investigated with numerical simulations \cite{Dalal:2007cu,Desjacques:2008vf,Pillepich:2008ka,Grossi_etal2009,GiannantonioPorciani2010,HamausSeljakDesjacques2011,Scoccimarro:2011pz,Biagetti:2016ywx}.   Since the inflationary consistency relation implies that no scale-dependent halo bias is generated in single field inflation \cite{Maldacena:2002vr,Tanaka:2011aj,Creminelli:2011rh,Baldauf:2011bh,Dai:2015jaa,dePutter:2015vga},  its detection in the bias of tracers on large scales (modulo projection effects \cite{Yoo:2010ni,Bonvin:2011bg,Challinor:2011bk,Jeong:2011as}) offers a means to rule out single-field inflation. This unique feature has been applied to constrain local PNG using galaxy survey data \cite{Slosar:2008hx,Afshordi:2008ru,Xia_etal2011,RossPercival_etal2013,Giannantonio:2013uqa,Leistedt:2014zqa,Ho:2013lda}. Although contaminations from late-time non-Gaussianity are relatively mild in the linear regime, it has been realized that the low-$k$ part of the power spectrum is susceptible to observational systematics, such as stellar contamination~\cite{RossPercival_etal2013,PullenHirata2013, Huterer_etal2013}. After carefully eliminating systematics, the current bound on $\fnl$ is $5 \pm 21$ (cross correlation between various data sets \cite{Giannantonio:2013uqa}) and $ -39 < \fnl < 23$ (quasars \cite{Leistedt:2014zqa}). The constraints from future surveys are expected to tighten by one to two orders of magnitude~\cite{HamausSeljakDesjacques2011,Dore:2014cca,Alonso:2015sfa,dePutter:2014lna,Raccanelli:2014kga, Yamauchi:2014ioa,Ferraro:2014jba,Ferramacho:2014pua}.  This can be achieved by combining multiple tracers of the LSS, which allows one to cancel out the dominant cosmic variance contribution on large scales~\cite{McDonald:2008sh,Seljak:2008xr,HamausSeljakDesjacques2011,HammausSeljakDesjacques2012,Abramo:2013awa}.

Almost all studies on PNG focus on tracers with positive bias parameters (except~\cite{SekiguchiYokoyama2012,Uhlemann:2017tex,Castorina:2018zfk}), with galaxies as the prime example. Voids are distinct from halos, because their large-scale bias ranges from positive to negative values as the void size increases~\cite{Hamaus:2013qja,Chan:2014qka}. In recent years, various clustering properties of voids have been measured using galaxy samples: redshift-space distortions around voids~\cite{Paz:2013sza,Hamaus:2015yza,Cai:2016jek,Hawken:2016qcy,Hamaus:2016wka,Achitouv:2016mbn,Hamaus:2017dwj}, the configuration-space void bias~\cite{Clampitt:2015jra}, the tracer bias around voids \cite{Pollina:2018ekp}, and the Baryon Acoustic Oscillations from voids~\cite{Kitaura:2015ubm}.   Because the clustering of voids enables us to probe a range of bias that is not accessible to halos, in this work we investigate void bias in the presence of local PNG and its potential constraining power on $\fnl$.  We consider voids defined both in dark matter, as well as halo density fields in real space.

\section{Theory}
We derive the void bias in PNG using the peak-background split formalism \cite{Kaiser1984,Mo:1995cs,ShethTormen1999,Desjacques:2016bnm}. To do so we consider the response of the void size distribution to long wavelength perturbations. The void size distribution can be modeled as a first-crossing distribution problem in the excursion set formalism~\cite{BCEK1991}.  To be concrete, let us consider a simple first crossing distribution  $\mathcal{F}$ for voids~\cite{Sheth:2003py}:
\beq
\label{eq:Fapprox}
 \mathcal{F}  ( \nu, \dv, \dc ) 
= \sqrt{ \frac{2  }{ \pi } } \exp \Big( - \frac{ \nu^2 }{ 2 }  \Big)  \exp\Big(  - \frac{| \delta_{\rm v} | }{\delta_{\rm c}  } \frac{ \mathcal{D}^2 }{ 4 \nu^2 }  - 2 \frac{\mathcal{D}^4  }{\nu^4}  \Big).
\eeq
The peak significance of a void is defined as  $ \nu \equiv  | \dv | / \sigma_{R_{\rm L} } $, where $\dv$  is void formation threshold (in the spherical collapse model the linearly extrapolated value is -2.72 \cite{Blumenthal_etal1992}) and $ \sigma_{R_{\rm L} } $ is the RMS of the density field evaluated using a top-hat filter of size $R_{\rm L}$, the Lagrangian size of the void. $\mathcal{D} \equiv | \delta_{\rm v}| / (  \delta_{\rm c} + | \delta_{\rm v} | ) $ denotes the so-called void-and-cloud parameter with  $  \delta_{\rm c} $ being the  halo collapse threshold (1.68 in the spherical collapse model \cite{GunnGott1972}). $\mathcal{F}$ (modulo a Jacobian) is the fraction of Lagrangian space volume characterized by $R_{\rm L}$ that will turn into voids of size $R_{\rm v}$ in Eulerian space. We can map the Eulerian void size to the Lagrangian one using the spherical collapse relation in~\cite{Bernardeau:1993ac}.  The void size distribution can then be written as
\beq
\label{eq:nv}
 n_{\rm v}  = \frac{1}{V_{\rm L}} \frac{d \ln \nu }{d \ln R_{\rm L}} \nu \mathcal{F} (\nu) , 
\eeq
where the Lagrangian void volume is $ V_{\rm L} = (4 \pi /3  ) R_{\rm L}^3 $.

To derive the effect of a long density fluctuation mode on small-scale ones in the local PNG case, we first split the {\it Gaussian} potential $\phi$ into long- and short-wavelength perturbations $ \phi = \phi_{\rm l} + \phi_{\rm s}$.  The  small-scale Bardeen potential $ \Phi_{\rm s}$ becomes $ \Phi_{\rm s} \approx  \phi_{\rm s} +  2 \fnl \phi_{\rm l} \phi_{\rm s}$.  The small-scale overdensity is obtained via the Poisson equation
\beq
\label{eq:deltas_modulated} 
\delta_{\rm s }(\bm{k})  = \mathcal{M}(k)   \phi_{\rm s}( \bm{k} ) (  1 +  2 \fnl \phi_{\rm l} ) . 
\eeq
The factor $ \mathcal{M}$ reads 
\beq
\mathcal{M}(k) = \frac{ 2 }{ 3  }\frac{ k^2 T(k) D(z) }{ \Omega_{m} H_{0}^2 },  
\eeq
where $ \Omega_{\rm m} $ and $H_0$ are the matter density and Hubble parameter at present time, $T(k) $ is the transfer function, and $D(z)$ the growth factor normalized to the scale factor in the matter-dominated era.    By considering the response of $n_{\rm v}$ to a long-wavelength perturbation, we obtain the linear Gaussian  bias from Eq.~\eqref{eq:Fapprox} as \cite{Sheth:2003py,Chan:2014qka} 
 \beq
 \label{eq:bG_PBS}
  b_{\rm v}^{\rm G} = 1 +  \frac{ \nu^2 -1 }{ \dv } + \frac{ \dv \mathcal{D} }{ 4 \dc^2 \nu^2  }.  
\eeq
In addition to $b_{\rm v}^{\rm G}$, there is an extra contribution $b_{\rm v}^{\rm NG} $ in the presence of local PNG. Following~\cite{Slosar:2008hx}, in the local PNG model a long-wavelength perturbation effectively rescales the amplitude of the small-scale fluctuations [Eq.~\eqref{eq:deltas_modulated}]. The amplitude of linear fluctuations is parametrized by $\sigma_8$, defined as the RMS of the density field at scales of $8\MpcOh$.  We can write the PNG bias as a response of $n_{\rm v } $ to the local value of $ \sigma_8 $ as~\cite{Slosar:2008hx}
\begin{align}
  \label{eq:bNG_dlndsigma8} 
  b_{\rm v}^{\rm NG}   & =  2 \fnl \mathcal{M}^{-1} \frac{ \partial \ln n_{\rm v}  }{\partial \ln \sigma_8(\bm{x} ) }.   
\end{align} 
For the case of halos one typically assumes the universality of the halo mass function and replaces the local $\sigma_8 $ amplitude by $\sigma_{ R_{\rm L} } $.  Following the same approach for voids, with Eq.~\eqref{eq:Fapprox} we obtain 
 \beq
 \label{eq:bv_PNG}
b_{\rm v}^{\rm NG}( k )  =   \frac{ 3 \fnl \Omega_{\rm m} H_0^2   }{ k^2  T(k)   D(z) } \bigg( \nu^2 - 1 - \frac{ |\dv | \mathcal{D}^2 }{ 2 \dc \nu^2 }   -   \frac{ 8 \mathcal{D}^4 }{\nu^4 }    \bigg).  
\eeq
This is similar to the PNG void bias derived in the high-peak limit derived in \cite{SekiguchiYokoyama2012}:
 \beq
 \label{eq:bv_NG_standard_highpk} 
b^{\rm NG}_{\rm v} = \frac{ 3 \fnl \Omega_{\rm m} H_0^2  }{ k^2 T(k) D(z)  } \delta_{\rm v} (b_{\rm v}^{\rm G} - 1).
 \eeq
It is the analog of the well-known PNG halo bias \cite{Dalal:2007cu,Matarrese:2008nc,Slosar:2008hx} (with $ \delta_{\rm v}$ replaced by $\delta_{\rm c} $) and agrees with Eq.~\eqref{eq:bv_PNG} in the high-peak limit, which is valid for large void sizes.

\section{Numerical results} 

The $N$-body simulation used in this work contains $1536^3$ particles in a box of 2000 $\MpcOh$ side length. There are three sets of different initial conditions: $\fnl=0$, $ 250 $, and $ - 250 $, with each eight realizations. The cosmological parameters are $\Omega_{\rm m} = 0.3$, $\Omega_{\Lambda}=0.7$, $n_s=0.967$, and $\sigma_8=0.85$. To compute $\partial \ln n_{\rm v} / \partial \ln \sigma_8 $ numerically,  we use two sets of Gaussian simulations with $\sigma_8=0.83$ and $0.87$. They share the same cosmological parameters as the other Gaussian simulations except $\sigma_8$ and there are two realizations each. The initial particle displacements are implemented using {\tt 2LPTic}~\cite{CroccePeublasetal2006,Scoccimarro:2011pz} at $z=99$, and then evolved with { \tt Gadget2} \cite{Gadget2}. Halos are identified using the halo finder {\tt Rockstar}~\cite{Behroozi_etal2013}. For more details, we refer the readers  to~\cite{Biagetti:2016ywx}.  Void catalogs are extracted using the void finder {\tt VIDE}~\cite{Sutter:2014haa}, which is based on {\tt ZOBOV}~\cite{Neyrinck2008} using a watershed algorithm~\cite{Platenetal2007}.

As a consequence of the typically large extent of voids, their number density is generally low and shot noise can be substantial. Furthermore, exclusion effects are significant and cause strong scale dependence on relatively large scales in the void auto-power spectrum~\cite{Hamaus:2013qja,Chan:2014qka}. Thus, the scale-dependent bias from PNG is most apparent when voids are cross-correlated with other tracer species. The cross-power spectrum between the species $i$ and $j$  is given by
\beq
\label{eq:P_ij} 
\langle \delta_i( \bm{k} ) \delta_j( \bm{k}' )  \rangle  =   P_{ij}( k ) \Ddel( \bm{k} +  \bm{k}' ) ,
\eeq 
where $ \delta_i$ and  $\delta_j$ are their overdensities, and $ \Ddel $ is the Dirac delta function. On large scales it can be expressed as $P_{ij} = b_ib_jP_{\rm  mm }+\mathcal{E}_{ij}$, where $b_i$ are the linear bias parameters (Gaussian or non-Gaussian), $P_{\rm  mm }$ is the matter power spectrum, and $\mathcal{E}_{ij}$ the shot noise matrix \cite{Hamausetal2010}.

\begin{figure}[!tb]
\centering
\includegraphics[width=\linewidth, trim = 10 20 10 36]{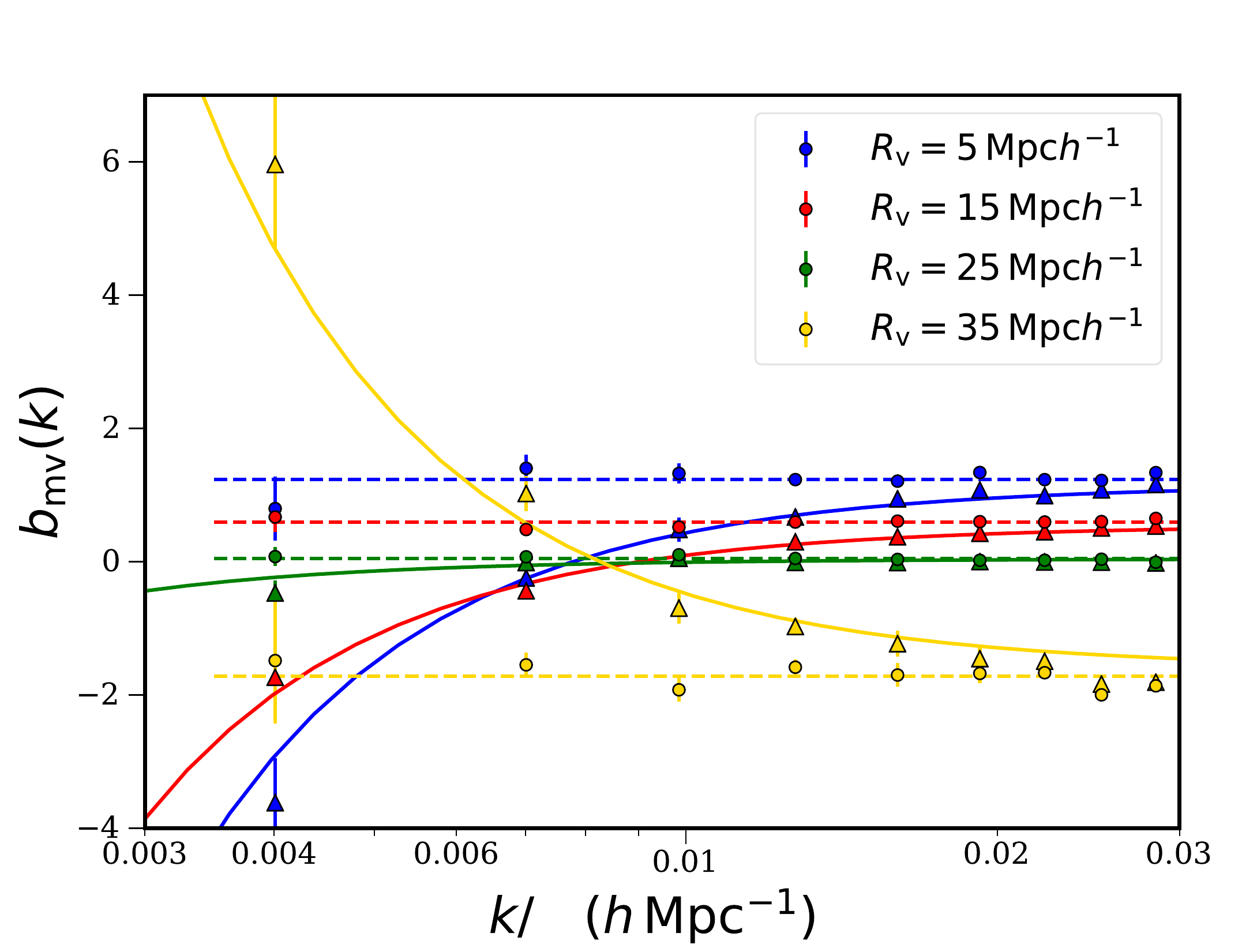} 
\caption{ Cross bias $b_{\rm mv }$ obtained using the cross-power spectrum between matter and voids at $z$=0. The measurement from the Gaussian (circles) and the PNG simulations with $\fnl=250$ (triangles) are shown for different void sizes (as labeled in the figure legend). The horizontal dashed lines are fit to the low-$k$ part of the Gaussian results, and the solid curves are the predictions obtained using Eq.~\eqref{eq:bNG_dlndsigma8}. }
\label{fig:bmv_PBS_fit_z0_subset_single}
\end{figure}

Fig.~\ref{fig:bmv_PBS_fit_z0_subset_single} shows the cross bias between matter and voids,  $b_{\rm mv } \equiv   P_{\rm mv } / P_{\rm  mm }$, both in the Gaussian and the PNG simulations ($\fnl=250$) for a range of void sizes at $z=0$. These voids are identified in the matter density field in real space with a sampling density of $0.005 \, (\hOMpc)^{3}$ and binned into different effective radii $R_{\rm v}$ of bin-width $10\MpcOh$. The Gaussian void bias becomes independent of scale at low~$k$, so we fit it with a constant up to  $k < 0.03 \hOMpc $. It decreases from  $b_{\rm v}^{\rm G} \sim 1 $ to 0  when the void size reaches $R_{\rm v}\sim 25 \MpcOh $, and goes negative when the void size further increases. The precise values of the bias parameters depend on the sampling density (c.f. \cite{Chan:2014qka}).  As shown in~\cite{Chan:2014qka}, the Gaussian bias from Eq.~\eqref{eq:bG_PBS} can qualitatively describe the trend in the simulations provided  that  $\dv $ is chosen less negative than its spherical collapse value of $-2.72$., e.g.~$\dv \simeq-1.0$ in~\cite{Chan:2014qka} (see also \cite{Pisani:2015jha}).

Furthermore, the PNG cross bias exhibits strong scale-dependence in the low-$k$ regime. We over-plot the prediction from Eq.~\eqref{eq:bNG_dlndsigma8} and find an excellent agreement with the numerical measurement. The derivative $\partial \ln n_{\rm v} / \partial \ln \sigma_8 $ is computed by finite differencing $n_{\rm v} $ measured in the Gaussian simulations with $\sigma_8=0.83$ and $0.87$. From Fig.~\ref{fig:bmv_PBS_fit_z0_subset_single} it is evident that the sign of the additional PNG bias changes from negative, when $b_{\rm v}^{\rm G} \gtrsim 0 $, to positive, when  $b_{\rm v}^{\rm G} \lesssim 0 $. Contrary to \cite{Castorina:2018zfk}, where zero-bias tracers are suggested to be very sensitive to PNG, our numerical results suggest the PNG signal to almost vanish for this type of tracer.

\begin{figure}[!tb]
\centering
\includegraphics[width=\linewidth, trim = 10 28 10 36]{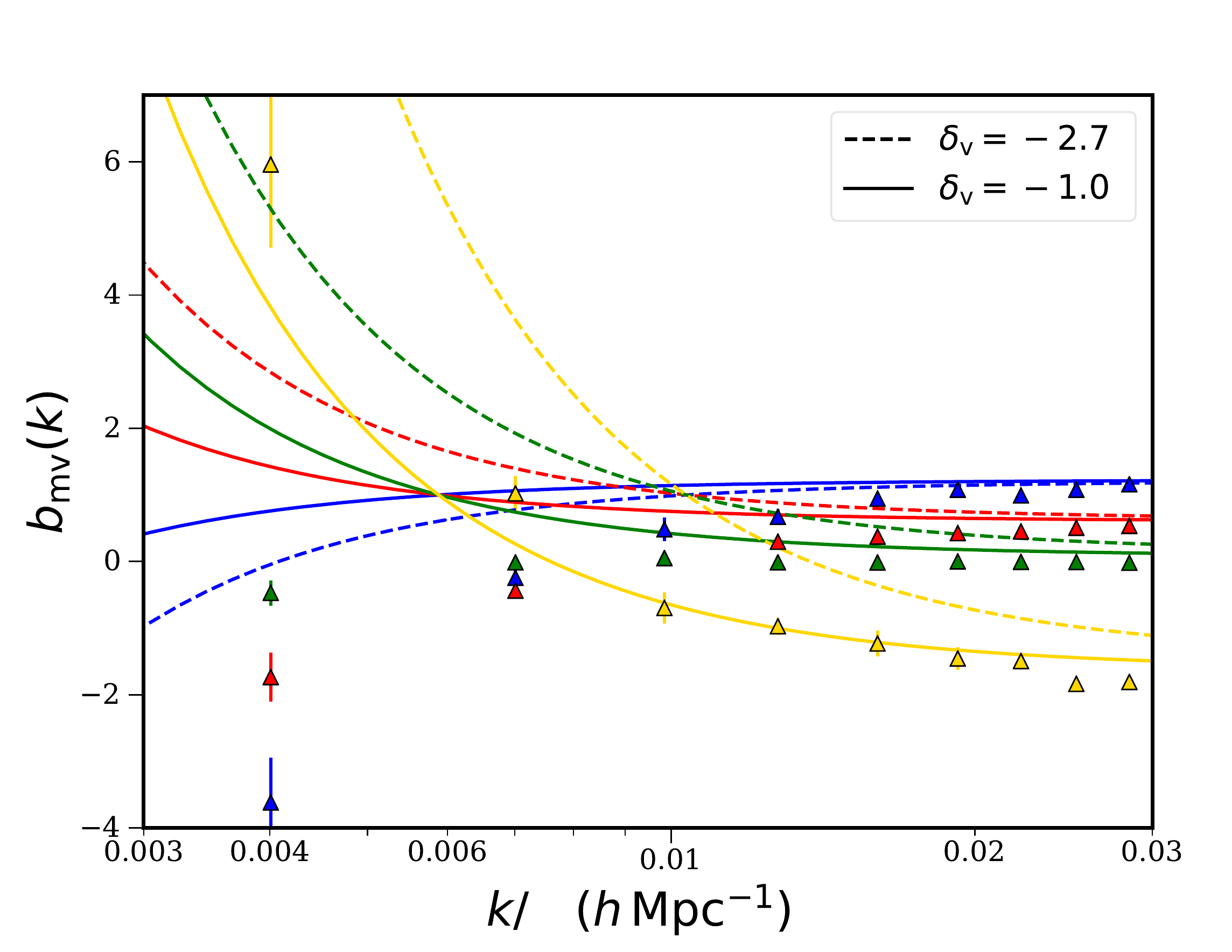}
\caption{ Comparison of the prediction for $b_{\rm mv }$ using Eq.~\eqref{eq:bv_NG_standard_highpk} for $\fnl=250$ with the numerical results (data points are the same as in Fig.~\ref{fig:bmv_PBS_fit_z0_subset_single}). For each void size bin, two values of $\delta_{\rm v}$ are shown: $-2.7$ (dashed) and $-1.0$ (solid). }
\label{fig:bmv_z0_DMVoid_b1naive_simple}
\end{figure}

We go on to test Eq.~\eqref{eq:bv_NG_standard_highpk}, which is expected to be a good approximation for large voids. In Fig.~\ref{fig:bmv_z0_DMVoid_b1naive_simple} we present the results for $\delta_{\rm v} $ = $-2.7$ and $-1.0$, which both fail to accurately describe the numerical results. Poor agreement is also found using Eq.~\eqref{eq:bv_PNG}.  Eq.~\eqref{eq:bv_NG_standard_highpk} implies that the PNG signal should flip sign at $b_{\rm v}^{\rm G} =1 $ instead of $b_{\rm v}^{\rm G} \sim 0  $.  If voids tend to remain at their Lagrangian position, then  $b_{\rm v}^{\rm G} $ is simply equal to the Lagrangian bias and the sign indeed flips at  $b_{\rm v}^{\rm G} =0 $, but recent work~\cite{Massara:2018dqb} shows that voids do move along with the LSS. However, to arrive at Eq.~\eqref{eq:bv_PNG} [or \eqref{eq:bv_NG_standard_highpk}], we implicitly replace  $ \sigma_{\rm 8} $ by $\sigma_{R_{ \rm L}} $ in the definition of $\nu$. These $\sigma$'s are evaluated at different scales. While interchanging them appears to be fine for halos, it leads to erroneous predictions for voids. Hence, for voids one should apply the more general peak-background split result of Eq.~\eqref{eq:bNG_dlndsigma8}. A further advantage of Eq.~\eqref{eq:bNG_dlndsigma8} is that it is independent of the definition criterion for voids, because it only depends on the measured void size distribution.

\begin{figure}[!htb]
\centering
\includegraphics[width=\linewidth, trim = 10 20 10 20]{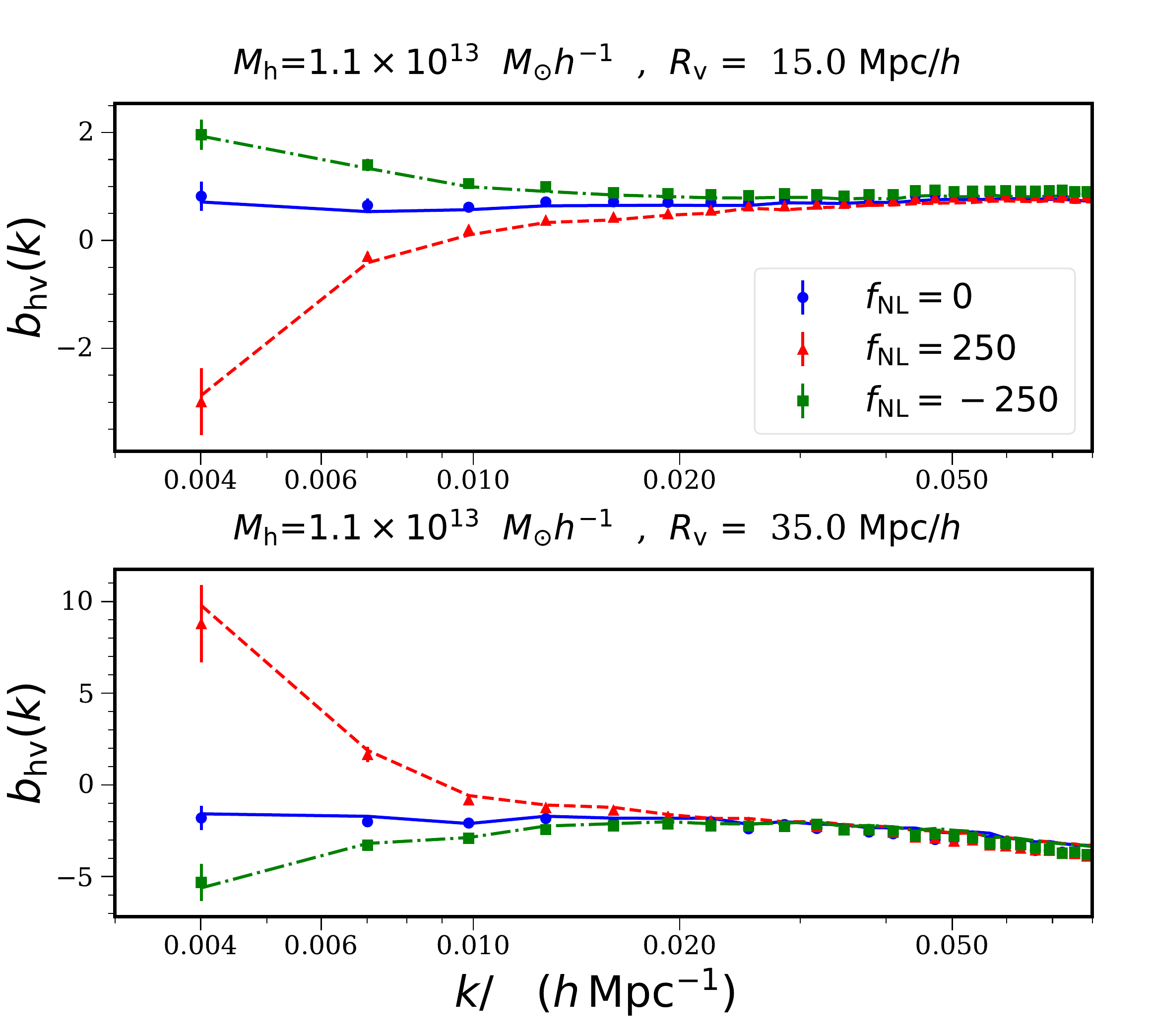} 
\caption{Cross bias $b_{\rm hv }$ between halos and voids at $z=0$. Measurements from Gaussian (blue circles) and PNG initial conditions with $\fnl=-250$ (green squares) and $\fnl=250$ (red triangles) are shown in comparison with $ b_{\rm mh} b_{\rm mv}  $ (blue solid, green dotted-dashed, and red dashed, respectively). Mean halo masses and void sizes are labeled above each panel.}
\label{fig:bhv_z0_smallsubset}
\end{figure}

In order to circumvent the difficulty in observing $P_{\rm mv}$ in galaxy surveys, we consider the cross-power spectrum between halos and voids, $P_{\rm hv}$. We define $b_{\rm hv} \equiv  P_{\rm hv} / P_{\rm mm}$. On large scales, where linear bias is valid, it is natural to expect 
$b_{\rm hv} \approx    P_{\rm mh} P_{\rm mv} / P^2_{\rm mm }$. In Fig.~\ref{fig:bhv_z0_smallsubset} we plot the numerical measurements of  $b_{\rm hv}$  between a halo bin of mean mass $M_{\rm h}= 1.1 \times 10^{13} \Msun $ and two different void-size bins, $R_{\rm v}=$ 15 and 35 $\MpcOh$. These void samples are the same as the ones used in Fig.~\ref{fig:bmv_PBS_fit_z0_subset_single}. There is significant scale dependence in the PNG case relative to the Gaussian one on large scales.  We also show  $ P_{\rm mh} P_{\rm mv} / P^2_{\rm mm }$ using the numerical power spectra. Indeed $b_{\rm hv} $ agrees well with $  b_{\rm mh} b_{\rm mv} $, but on smaller scales where nonlinearity and nonlinear biasing kick in, we expect deviations from this simple relation (although it is not apparent for this halo mass bin).

\section{Fisher forecast}

In this section we study how much additional information on $\fnl$ can be gained by including voids in a multitracer analysis using the Fisher matrix formalism (see \cite{Heavens2009,Tegmark:1996bz,Albrecht:2006um} for a review). For simplicity, we consider $\fnl$ as the only free parameter, as degeneracies with other cosmological parameters are expected to be of minor importance when multiple tracers of the same underlying density field are considered~\cite{McDonald:2008sh,Seljak:2008xr,HamausSeljakDesjacques2011}. Assuming the Fourier modes of the overdensities of halos and voids are Gaussian distributed, the Fisher matrix of the fields reads
\begin{align}
  \label{eq:Fisher_Pk_GaussL} 
  F_{ \fnl  \fnl} =  V \int  \frac{ d^3 k  }{ (2 \pi)^3 } \frac{1 }{ 2 } \mathrm{Tr}\Big( \Sigma^{-1} \frac{ \partial \Sigma  }{ \partial \fnl }  \Sigma^{-1} \frac{ \partial \Sigma  }{ \partial \fnl }    \Big) , 
\end{align}
where $ V  $ is the survey volume and $ \Sigma (k) $ the covariance matrix of the multitracer density field~\cite{HamausSeljakDesjacques2011}. The elements $\Sigma_{ij} (k) $ are simply given by all possible auto- and cross-power spectra $P_{ij}(k)$ from Eq.~\eqref{eq:P_ij}.  We consider a survey volume of 8$\,\mathrm{Gpc}^3h^{-3}$ at $z=1$, and compare constraints on $\fnl$ from a multitracer survey of halos, voids, and a combination of the two. To be more realistic, voids are constructed from the distribution of halos in real space rather than the dark matter. In each case, the multitracer technique takes advantage of sampling variance cancelation in the primordial modes, which are fairly uncorrelated on large scales~\cite{Seljak:2008xr,HamausSeljakDesjacques2011}. 

In calculating $ F_{ \fnl  \fnl}$, we directly use the numerical power spectra, including their shot noise contribution. The $\fnl$-response derivative  $\partial \Sigma /  \partial \fnl $ is obtained via finite differencing of the $\fnl=250 $ and $\fnl=-250  $ power spectrum measurements. We remove the scale-independent contribution due to differences in shot noise between the catalogs. To increase the accessible range in number densities, we use halos containing at least five particles. As we are primarily interested in the large-scale modes of the density field, the fact that these halos are not well resolved is of minor importance.  Using higher-resolution simulations we have verified that our forecast is robust up to an uncertainty of at most $10\%$\footnote{We used the Carmen simulation from the LasDamas project (\url{http://lss.phy.vanderbilt.edu/lasdamas/overview.html}) to perform a partial resolution study. The LasDamas simulations are Gaussian, so we rely on the $\fnl$-response measured at lower resolution, but obtain  more accurate power spectra.}. We divide the full halo sample (minimum 5 particles) into five mass bins, with mean halo masses  $M_{\rm h}=1.33 \times 10^{12}$,  $4.86 \times 10^{12} $, $1.10 \times 10^{13} $, $3.34 \times 10^{13} $, and $1.44 \times 10^{14} \, \Msun $. Voids are extracted from the halo samples with various mass thresholds (minimum 5, 10, 20, or 50 particles). The resulting void samples are further divided into three bins of void size $R_{\rm v}=$ [0, 20], [20, 40], and [40, 80] $\MpcOh$.  A detailed modeling of the PNG signals from these void samples is beyond the scope of this paper and will be investigated elsewhere.

\begin{figure}[!tb]
	\centering
	\includegraphics[width=\linewidth, trim = 10 0 10 -17]{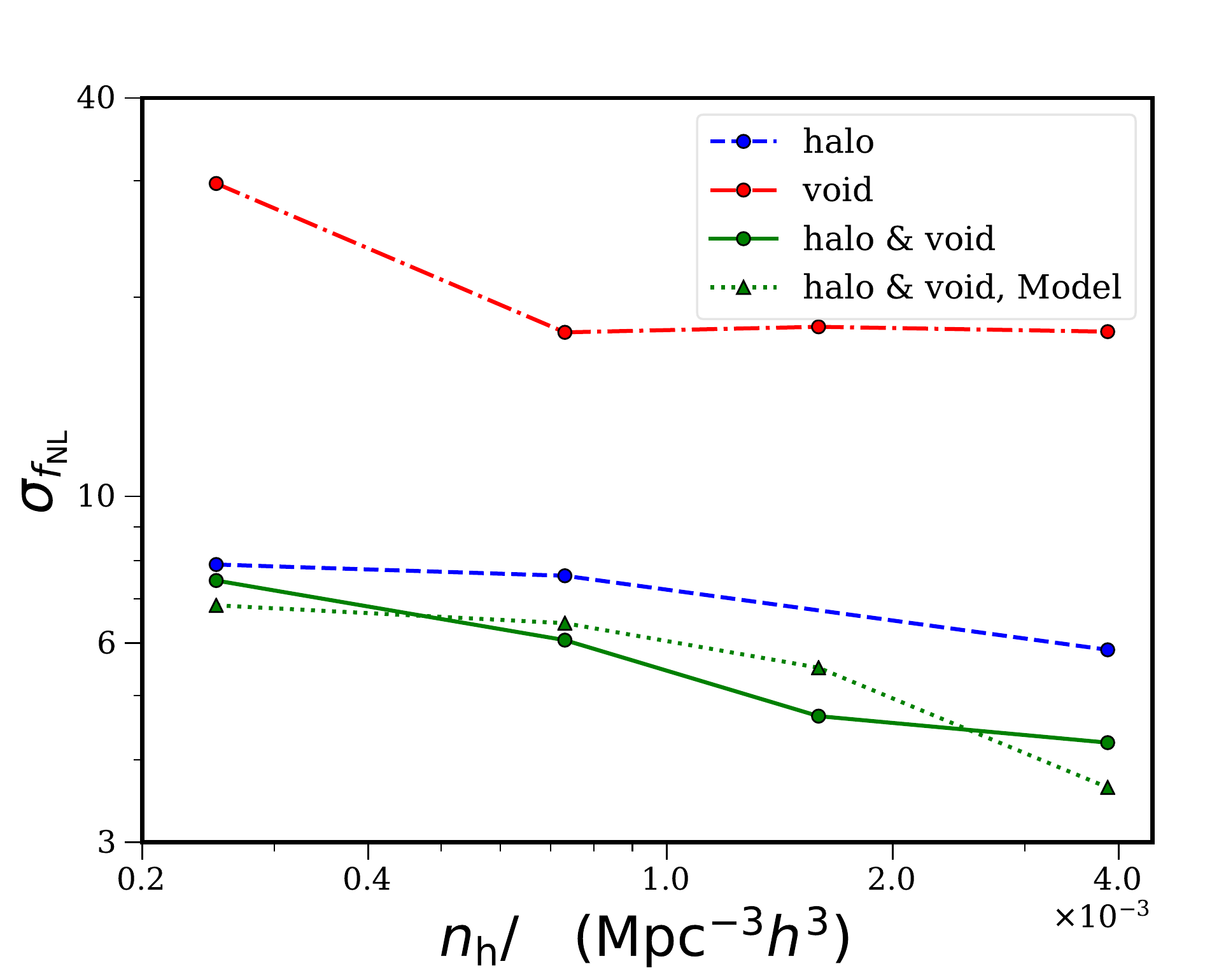}
	\caption{Multitracer Fisher forecast on $\sigma_{\fnl}$ for different tracers up to $ k_{\rm max} =0.08 \hOMpc $ at $z=1$ (blue dashed for halos, red dot-dashed for voids, and green solid for both). The green dotted line is the theoretical expectation from Eq.~(29) of \cite{HamausSeljakDesjacques2011} for the combined case. The halo number densities $n_{\rm h}$ shown on the $x$-axis are used to construct the void samples.}
	\label{fig:sigma_fNL}
\end{figure}

Fig.~\ref{fig:sigma_fNL} presents the constraints on $\fnl$ via $\sigma_{\fnl} \equiv 1/ \sqrt{ F_{\fnl \fnl} } $, using all Fourier modes up to $k_{\rm max} = 0.08 \hOMpc$. The increasing number densities correspond to halo samples with at minimum 5, 10, 20, or 50 particles, respectively. The halo constraint is obtained using the available mass bins depending on the minimum particle threshold\footnote{Because the lowest halo mass bin corresponds to 5-20 particles, the halo constraint for minimum 10-particles is not shown. For the joint constraint in this case, we use all the halo mass bins. }. The constraints from voids alone are weak compared to those from halos due to their high shot noise, saturating for $n_{\rm h} \gtrsim 7 \times 10^{-4} \, \hOMpcCube $. When the number density of halos increases, the distribution of void sizes shifts to smaller voids. As we are keeping the bins of void sizes fixed, this means we are losing the largest voids that contribute the strongest signal on $\fnl$, due to their very negative bias amplitudes. However, when voids and halos are combined in a multitracer analysis, the joint constraints improve appreciably thanks to the large cross-correlations and the low shot noise between them. In particular, when the halo number density reaches $4 \times  10^{-3} \, \hOMpcCube  $, the error on $\fnl $ is reduced by almost a factor of 2 compared to the halo case. This result is in good agreement with the analytical Fisher forecast based on Eq.~(29) in \cite{HamausSeljakDesjacques2011}, shown as the green dotted line in Fig.~\ref{fig:sigma_fNL}. In that calculation, we used the $\fnl$ response from the cross-power spectra of each tracer with the matter field and assumed Poisson shot noise.

\section{Conclusion}
Using a suite of $N$-body simulations, we have demonstrated that in the local PNG model, voids exhibit a scale-dependent bias on large scales, just like halos. Although the standard calculation that is analogous to the halo case  [Eq.~\eqref{eq:bv_PNG} or \eqref{eq:bv_NG_standard_highpk}] fails to describe the simulation results for voids, the general peak-background split prediction [Eq.~\eqref{eq:bNG_dlndsigma8}] yields an excellent agreement. Furthermore, based on the Fisher matrix formalism for multiple tracers we have shown that by combining the clustering information from voids and halos, constraints on $\fnl$ can substantially be tightened, as long as the number density of tracers is sufficiently high. Our simplistic analysis using a volume of $8\,\mathrm{Gpc}^3h^{-3}$ and tracer densities up to $4\times10^{-3}\, \hOMpcCube $ already renders $\fnl$ constraints of $\mathcal{O}$(a few) achievable. Although we only show the results in real space, the effects from redshift-space distortions on the clustering of voids on large, linear scales is well studied \cite{Paz:2013sza,Hamaus:2015yza,Cai:2016jek,Hawken:2016qcy,Hamaus:2016wka,Achitouv:2016mbn,Hamaus:2017dwj,Chuang:2016wqb}.  Optimizing the binning strategy in constructing multiple tracers from halos and voids will most likely yield further gains~\cite{Hamausetal2010,HamausSeljakDesjacques2011}. Future surveys, such as Euclid, will have access to even larger volumes and higher densities of tracers, opening up the possibility to significantly improve upon current CMB constraints on PNG with the help of cosmic voids. The latter are contained in the survey data anyway and hence provide additional information at no cost.

\section*{Acknowledgements}
We thank Vincent Desjacques for sharing his numerical simulations with us for the initial phase of the study, and Neal Dalal for fruitful discussions. KCC acknowledges the support from the National Science Foundation of China under the grant 11873102. NH acknowledges support from the DFG cluster of excellence ``Origins'' and the Trans-Regional Collaborative Research Center TRR 33 ``The Dark Universe'' of the DFG. NH is grateful to the Institut Henry Poincar\'e for hospitality during the ``CosmoMath'' trimester in the fall 2018. MB acknowledges support from the Nederlands Organisation for Scientific Research (NWO) under the VENI  grant  016.Veni.192.210.

\bibliography{reference} 

\end{document}